\documentclass[11pt]{article}
\usepackage[utf8]{inputenc}
\usepackage{authblk}
\usepackage{amsmath,amsfonts,amssymb}
\usepackage{graphicx}
\usepackage[colorlinks=true, allcolors=blue]{hyperref}
\title{\textbf{Point defects in IBS coating for very low loss mirrors.
}}

\author[1,2,*]{Sihem Sayah}
\author[1]{Benoît Sassolas}
\author[1,2]{Jérôme Degallaix}
\author[1]{Laurent Pinard}
\author[1]{Christophe Michel}
\author[2]{Viola Sordini}
\author[3]{Gianpietro Cagnoli}
\affil[1]{Laboratoire des Matériaux Avancés-IP2I, CNRS/IN2P3, Univ Lyon, Villeurbanne, France}
\affil[2]{Univ Lyon, Université Claude Bernard Lyon 1, IP2I Lyon, CNRS, F-69622, Villeurbanne, France}
\affil[3]{Univ Lyon, Université Claude Bernard Lyon 1, Institut Lumière Matière, CNRS/IN2P3, Villeurbanne, France}
\affil[*]{Email: s.sayah@lma.in2p3.fr}

\date{} % comment this line to show today's date 

\providecommand{\keywords}[1]
{
  \small	
  \textbf{\textit{Keywords :}} #1
}

\begin{document}

\maketitle

\begin{abstract}
High reflective coatings are used in many physics experiments. 
Despite the high quality of the optical coating, the performances of the mirrors is altered by the scattered light induced by micrometers size defects in the coating layers. The topic of this paper is the study of the point-like scatterers present in the specific coating of the mirrors used in state of the art, high sensitivity optical experiments. We studied the behavior of the materials according to different thicknesses, and how the defects change after annealing. To our knowledge, this is a first insight into the formation of such defects for different materials and thickness and how this is reduced when samples are annealed.
   
\end{abstract}


\section{Introduction}

The year 2015 was marked by the first direct detection of a Gravitational Wave (GW), a confirmation of Albert Einstein’s theory of general relativity proposed a century ago \cite{GW}. This discovery offers us a new way to observe and understand the Universe around us. The detection was achieved by the two laser interferometers LIGO which were later joined by Virgo for subsequent detections \cite{O2,O3}. Such gravitational waves detectors are kilometers-long Michelson interferometers with Fabry-Perot cavities in the arms. At their heart, we find some of the most reflecting mirrors in the world. The Laboratoire des matériaux avancées (LMA) in Lyon (France) has performed the coating and metrology on the main mirrors of all the existing detectors \cite{jerome}.

The technology of such mirrors is also used for the most precise optical clocks \cite{clock} and in advanced physics experiments \cite{BMV,cavity}.

However, the mirrors exhibit point defects in the thin layers that scatter the light inducing a loss of the laser power of the order of a few tens of parts per million (ppm). Light loss has two consequences: a loss of optical power in the interferometer arm amplified by the arm cavity gain and the addition of a phase noise when the scattered light is recombined to the main beam after reflection on the walls of the vacuum tubes that are mechanically excited by the micro seism \cite{stat}.
Both of these phenomena limit the sensitivity of the detector impacting the ability to detect astrophysical events \cite{noise}. The scattered light is one of the main sources of noise under study for the Advanced Virgo upgrade scheduled for 2021 \cite{AdVirgo+}. The improvement of these detectors requires an understanding of the origin and nature of the defects in order to reduce their densities. This research effort is essential to improve the optical performances of the mirrors in view of the next generation of coating. This paper describes the first results of the analysis of the relation between the deposition parameters and the point-like scatterers density for tantala and silica thin films.

\section{Experimental set-up}

The mirrors in the interferometers are Bragg reflectors composed of stacks of thin layers, alternating thin films made of $\text{Ta}_{2}\text{O}_{5}$ and $\text{SiO}_{2}$. 

Each mirror was characterized after coating in terms of defect density and scattering level. These measurements provided information for the full stack independently of the layer material. In order to better understand the process generating the defects, we studied tantala and silica monolayers separately. Moreover several influence parameters were tested like the layer thickness and the annealing in order to determine their impact on the point-like scatterers density.

\subsection{Coating Process}

Deposition is performed at the Laboratoire des Matériaux Avancés (LMA) in Lyon using a Ion Beam Sputtering (IBS) process \cite{performancesMiroirs}. The IBS deposition is done in a vacuum chamber, with a 12 cm Veeco ion source providing a 200 mA Argon positive ion beam at 1 keV \cite{IBS}. We deposited layers with different thicknesses on micropolished fused silica substrates of 1'' diameter from Coastline Optics. These substrates are of very good quality and have a very low defect density of the order of 0.04 defects/mm$^2$ and a RMS roughness around 1 \AA \: RMS.

For this study, we coated 8 samples : 4 $\text{Ta}_{2}\text{O}_{5}$ samples and 4  $\text{SiO}_{2}$ samples, with four comparable values of thicknesses (the exact values change slightly depending on the material).

\subsection{Characterization and measurement}

This section describes the instruments and techniques used to characterise the samples. Each sample was characterized after coating in terms of defect density and scattering level.

\subsubsection*{Description of the instrument}

After the depositions, the samples were studied with an instrument called MICROMAP \cite{micromap}, an optical profilometer that has been customized for the detection of defects in dark field illumination. The MICROMAP scans surfaces and when it detects a contrast difference in the field, the image is stored. A post treatment measures the defect size. The instrument is able to detect point-like defects with a size within 1 and 5 microns, giving a cartography of the sample with the number, size and localisation of the point-like-scatterers on the sample. It scans an area of 18 mm diameter centred on the 1'' sample in order to prevent from counting defects near the edge of the coating.

\subsubsection*{Detection method}

Before starting the detection of defects measurements, there are 2 important settings to check: the threshold and the background level. The first one is used to fix a grey level value beyond each defects will be count and the second one is the background noise. To be able to compare the 2 materials, the same process was used for the measurements. First we checked the background noise of the samples of the different materials and then we set the threshold at 19 for all the samples.

For $\text{SiO}_{2}$ samples, this is a layer of $\text{SiO}_{2}$ on fused silica. It is expected that the contrast as a function of the film thickness will not change. While for $\text{Ta}_{2}\text{O}_{5}$ on fused silica, we expected an influence of the background of the image to change with the thickness of the layer. This can have an impact on what the system can detect.

In table \ref{tab:seuillage}, we present the adjustment of MICROMAP.
For the 4 samples of each materials we have the same level of background noise. So we set the same thresholds for $\text{Ta}_{2}\text{O}_{5}$ and $\text{SiO}_{2}$. We notice that there is no influence of the layer thickness on the background level. The average background noise level is the same for the 4 thicknesses of the same material. So we can conclude for dark field detection, the background level of the measurement is no impacting on the ability to detect defects. The only significant difference is about the level of brightness for the focus map: it is lower for the $\text{Ta}_{2}\text{O}_{5}$ because it reflects more than $\text{SiO}_{2}$.

\begin{table}[htbp]
\centering
\caption{\bf Contrast setting of the two materials acccording to the different thicknesses .}
\begin{tabular}{|l|cccc|}
\hline
 Material &  \; Thickness & Thresholds & Light level & Background \\
   &  \quad (nm) & (grey level) & (\%) & level \\\hline\hline
 %Substrate &  \: 0 & 0.04 & 0  & \\\hline
           &  \quad 523 & 19 & 27 & 16.2\\
 $\text{SiO}_{2}$ &  \: 1005 & 19 & 27 & 16.1\\
            & \: 1550 & 19 & 27 & 16.2\\
            & \: 2053 & 19 & 27 & 16.3\\\hline
            & \quad 543 & 19 & 17 & 16.4\\
 $\text{Ta}_{2}\text{O}_{5}$ & \: 1083 & 19 & 17 & 16.5 \\
           &  \: 1626 & 19 & 17 & 16.6\\
           &  \: 2167 & 19 & 17 & 16.6\\\hline
\end{tabular} 
  \label{tab:seuillage}
\end{table}

\subsubsection*{Scattering measurement}
The scattering light is measured by a C.A.S.I. Scatterometer (Complete Angle Scan Instrument) \cite{casi}. This instrument composed by a laser and a detector allows to measure the scattered light at different scattering angles. A sample is illuminated at variable incidences and the detector scans the plane to measure the Bidirectional Reflectance Distribution Function (BRDF). The integrated value considers a pure isotropic emission, then this integrated value is given in parts per millions (ppm). Measurements are made with a source laser at 1064 nm, the same wavelength used by LIGO and Virgo interferometers.

\section{Experiment}
\subsection{Influence of the thickness of the layer on the number of defects}

Different samples were produced with different thicknesses (see Table \ref{tab:shape-functions}). The layers were deposited under the same conditions.

\begin{figure*}[htbp]
\centering
\includegraphics[width=0.48\textwidth]{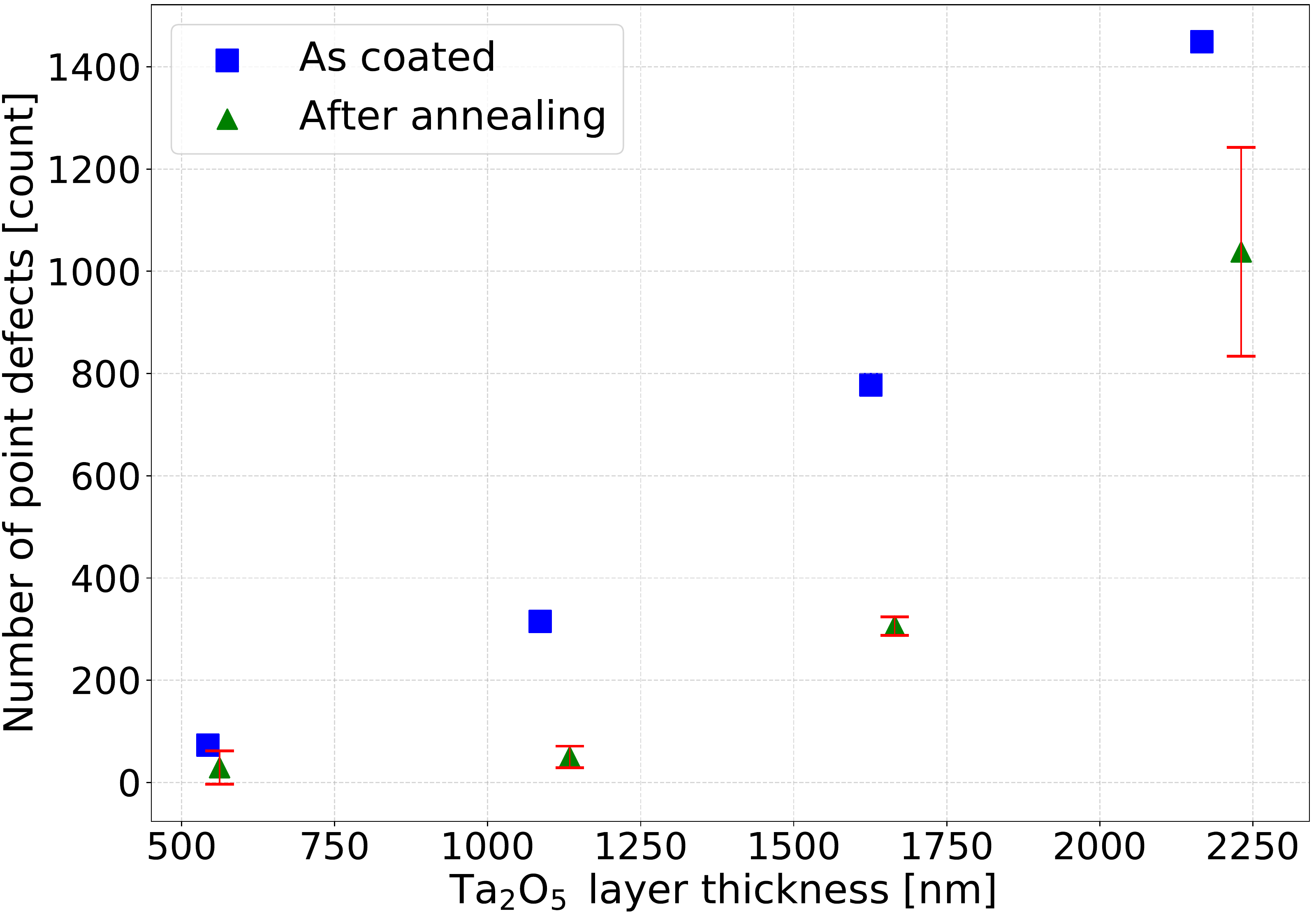} \hfill
\includegraphics[width=0.48\textwidth]{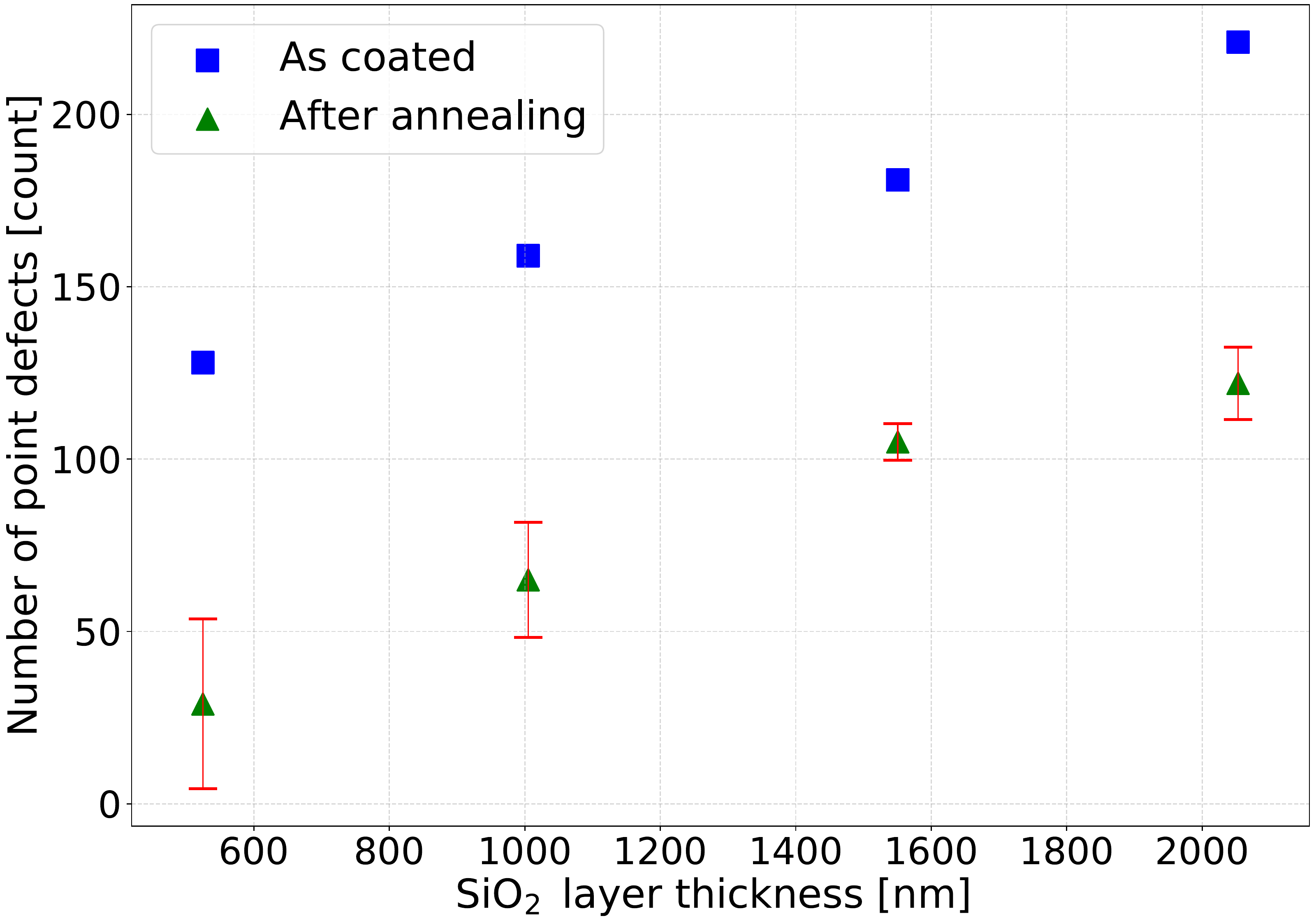}
\caption{Number of point-like scatterers in $\text{SiO}_{2}$ layers (right) and $\text{Ta}_{2}\text{O}_{5}$ (left) as a function of the layer thickness. Results are shown for materials as coated (in blue) and after annealing (in green). The measurements were repeated and the markers shows the average values, while the error bars show the region containing 90\% of the results. The actual value is the average of repated measurement. Measurements before annealing were not repeated. }
\label{fig:silice-tantale}
\end{figure*}

Figure \ref{fig:silice-tantale} shows the number of defects according to the layer thickness before and after annealing for $\text{SiO}_{2}$ (right) and $\text{Ta}_{2}\text{O}_{5}$ (left) respectively. We observe very precisely, for the two materials, that the number of point-like scatterers increases with the layer thickness.

\subsection{Influence of the annealing on the number of defects}

The impact of the post deposition annealing on the number of defects has been also studied. All samples were annealed in an oven in air at 500$^\circ$C for thermal annealing after deposition during approximatively 10 hours. 

For this, we measure layer thickness of each sample with a spectrophotometer. The refractive index and the thickness are first evaluated from the transmission spectrum using an extrema method \cite{enveloppe}. Then these results were used as starting values for numerical optimization between experimental and computed transmission values. For the $\text{SiO}_{2}$ layer, the refractive index contrast between the layer and the substrate is not enough to provide significant interferences pattern in the spectrum. For this purpose a SF11 substrate is coated at the same time as the micropolished fused silica substrate. Unfortunately after the annealing we have a delamination which makes the thickness measurement impossible after annealing.

A thickness increase with annealing has been observed previously, and it is more visible for $\text{Ta}_{2}\text{O}_{5}$ than $\text{SiO}_{2}$. As a reminder, in the present study thickness of $\text{SiO}_{2}$ samples were not measured after annealing. 

According to figure \ref{fig:silice-tantale}, annealing is observed to reduce the density of point defects by 40\% for silica layers and by 50\% for tantala. Nevertheless, the curves retain the trend observed before annealing.

\subsection{Study of defects size}

The MICROMAP software does not provide the size information of these defects, besides the fact that it is within 1 and 5 micrometers. The image stored by the MICROMAP are processed with an image-recognition software, to determine the radius of the detected defects (assumed to be circular), using the Laplacian of Gaussian algorithm in python as implemented in scikit-image library \cite{python}.

The LoG algorithm is based on a convolution of the image with 2-dimensional Gaussian functions with variable width. The defects are then detected looking at the local maxima of the
convoluted image, with amplitude above a given threshold. A calibration procedure has been developed, so that the threshold is chosen automatically based on the median brightness of each image. Before undergoing the defect detection step, the images stored by the MICROMAP are preprocessed with a low-pass filter in order to remove some background patterns.

\begin{figure*}[htbp]
\centering
\includegraphics[width=0.49\textwidth]{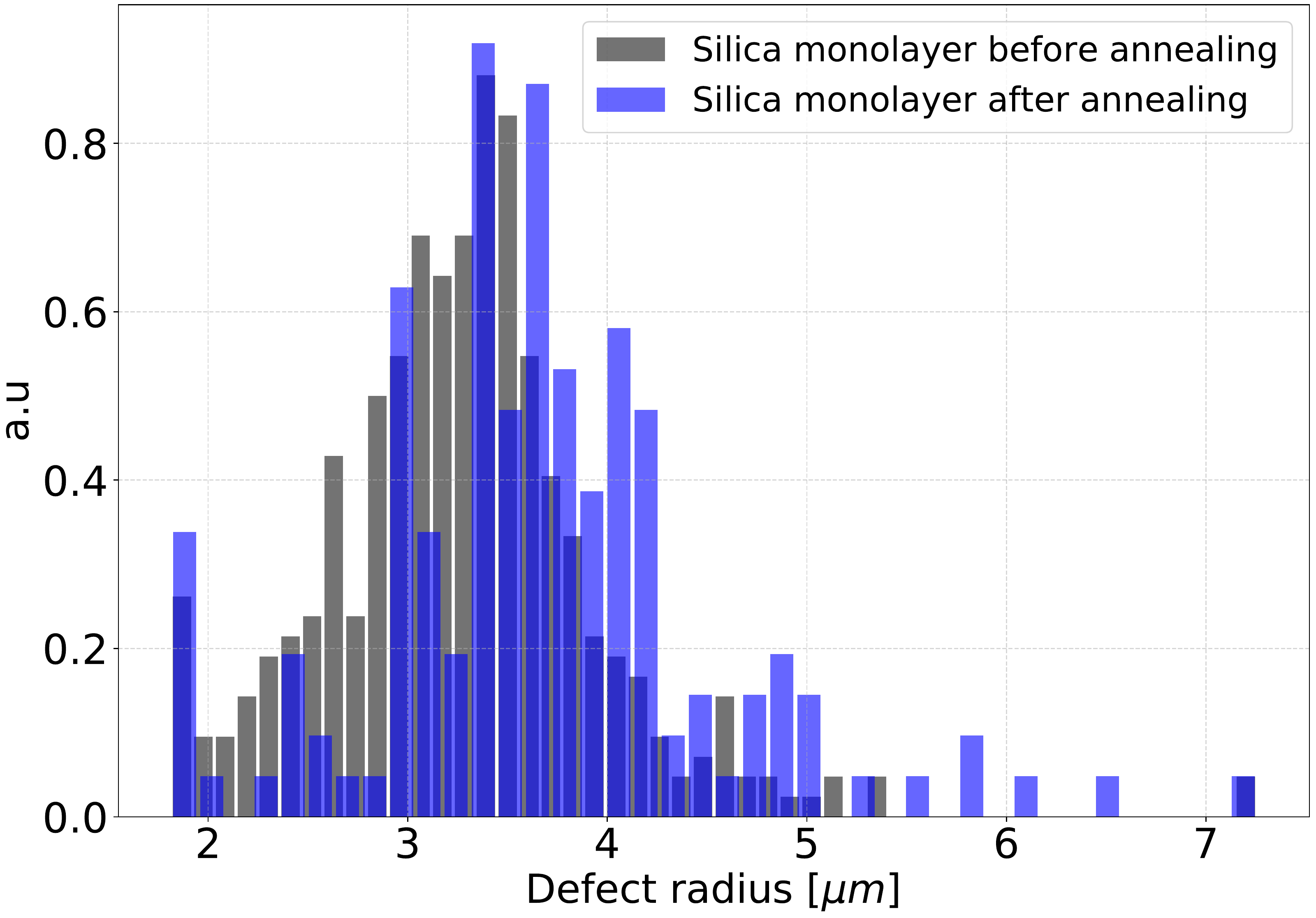} \hfill
\includegraphics[width=0.49\textwidth]{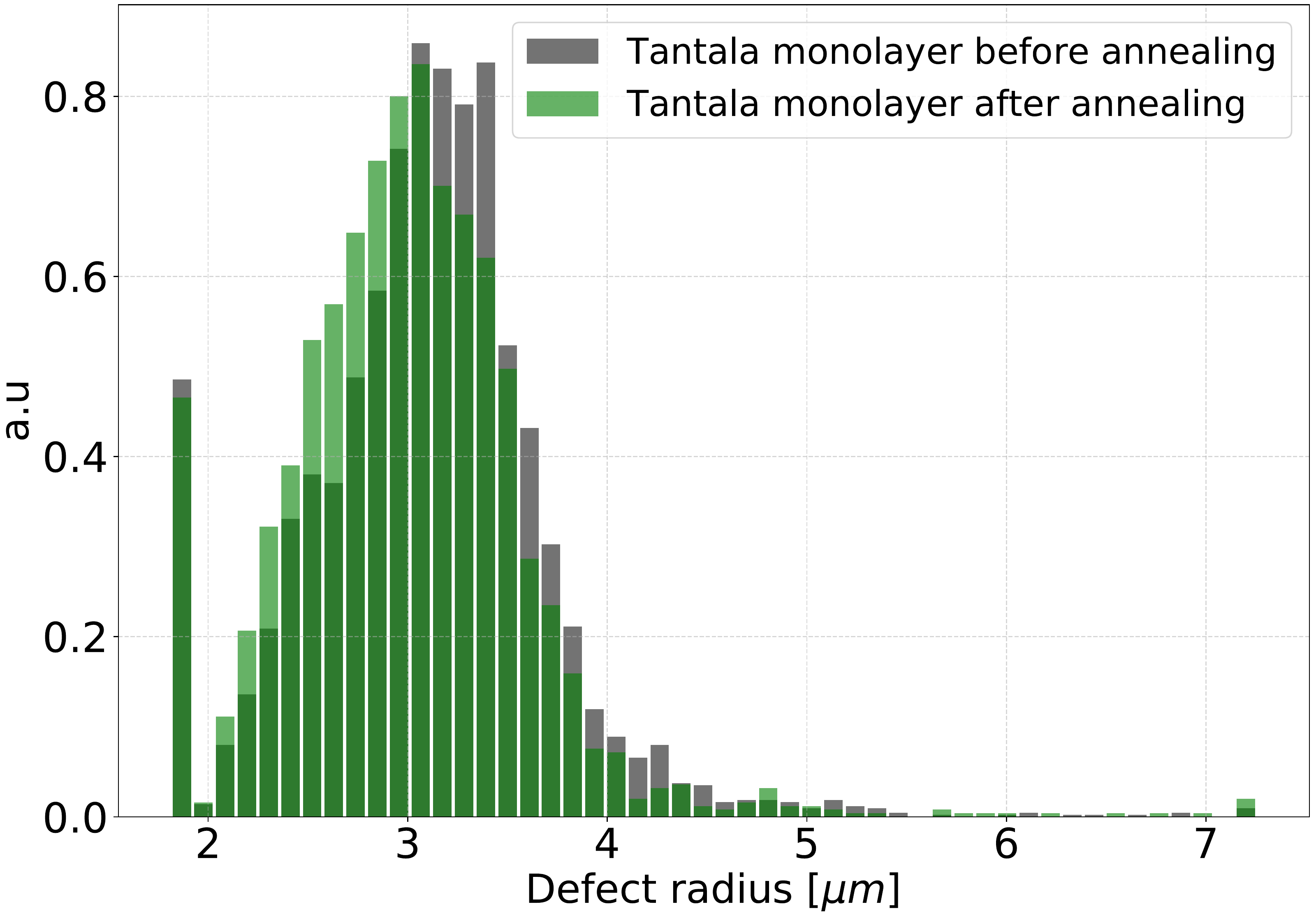}
\caption{Defect size for the thickest samples of the two materials, $\text{SiO}_{2}$ (left) and $\text{Ta}_{2}\text{O}_{5}$ (right), before and after annealing. The distributions have been normalised to ease the shape comparison. }
\label{fig:silice-tantale-histo}
\end{figure*}

Figure \ref{fig:silice-tantale-histo} shows the defects size for the thickest $\text{Ta}_{2}\text{O}_{5}$ and
$\text{SiO}_{2}$ samples, for which we have the highest number of defects.
The distributions have been normalised to ease the shape comparison.
After annealing, the median defects size is 3.66 $\mu$m for
$\text{SiO}_{2}$ and 2.96 $\mu$m for $\text{Ta}_{2}\text{O}_{5}$, with a 68\% probability region of
[3.04, 4.26] $\mu$m and [2.37, 3.48] $\mu$m, respectively, showing that the
defects size is lower for $\text{Ta}_{2}\text{O}_{5}$ than for $\text{SiO}_{2}$. The 95\% probability regions are quite large, because of the limited statistics, and do not allow to make quantitive comparisons. Also, the annealing seems to slightly lower the defects size for $\text{Ta}_{2}\text{O}_{5}$, while the opposite is observed for $\text{SiO}_{2}$.

\begin{figure}[htbp]
\centering
\includegraphics[width=0.5\textwidth]{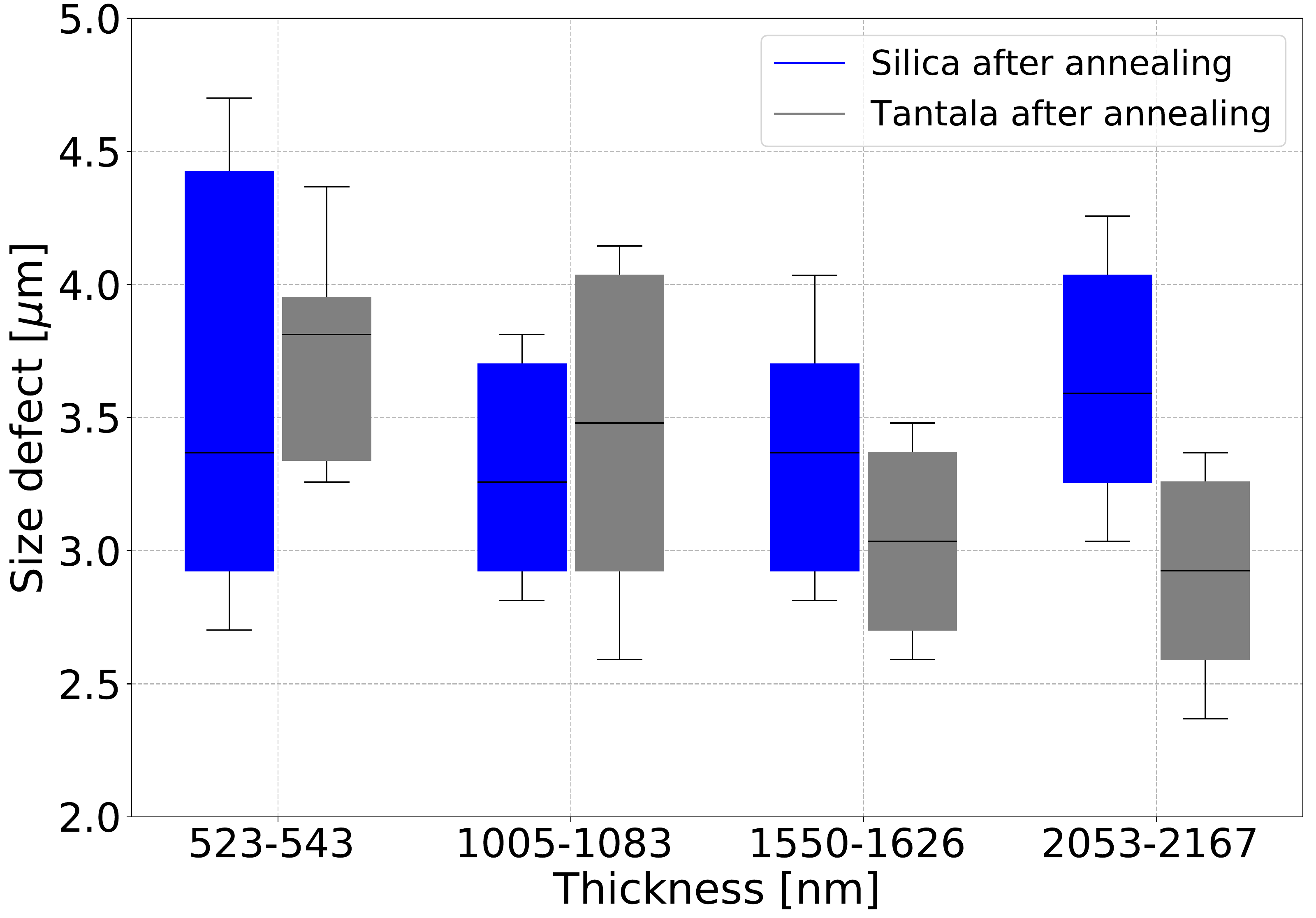}
\caption{Defect size for the two materials, $\text{Ta}_{2}\text{O}_{5}$ and $\text{SiO}_{2}$, after annealing and as a function of the sample thickness.  
The horizontal markers show the median, the colored bands and error bars  the 50\% and 68\% probability regions.}
\label{fig:moustache}
\end{figure}

Figure \ref{fig:moustache} shows the defect size median, 50\% and 68\% probability regions for the samples of four different thicknesses of $\text{SiO}_{2}$ and $\text{Ta}_{2}\text{O}_{5}$, after annealing. The low number of defects on the samples other than the thickest one affect the results, preventing us from drawing quantitative conclusions. 

A trend can be observed in the typical defect size in Tantala that seems to decrease with the sample thickness, while it remains constant for Silica. Although we cannot conclude that the defect size distribution is layer-dependent due to the wide dispersion which may not be significant. The need for statistics in order to study with precision the defect size is definitely a lesson learnt from this first measurements.

\subsection{Scattering light from defects points}

To quantify the impact of the defects on the scattered light, we measured the 8 samples with the CASI after annealing. Measurement were made on a diameter of 16 mm. The results are summarized in Table \ref{tab:shape-functions}. The thickest $\text{SiO}_{2}$  sample has a scattering of 11.5 ppm whereas the thinnest one 5.1 ppm. Similarly, for $\text{Ta}_{2}\text{O}_{5}$ samples, the thinnest sample scatters 5.8 ppm while the thickest, 11.6 ppm. 

It is worth mentioning that the uncoated substrates used in this study have an optical scattering level below 3 ppm. So the thin layer coating and particularly the addition of defects contribute to the scattered light.

While there is a factor 10 on the number of defects between the thickest sample of silica and tantala, the scattered light is comparable. This tends to show that the defects present in the layers of $\text{Ta}_{2}\text{O}_{5}$ scatter less light than the defects present in the $\text{SiO}_{2}$. In addition, the distribution of the electric field being different in the two materials could also affect the scattered light. This effect is not studied in detail here but will be taken into account in the future. 

\begin{figure*}[htbp]
\centering
\includegraphics[width=0.75\textwidth]{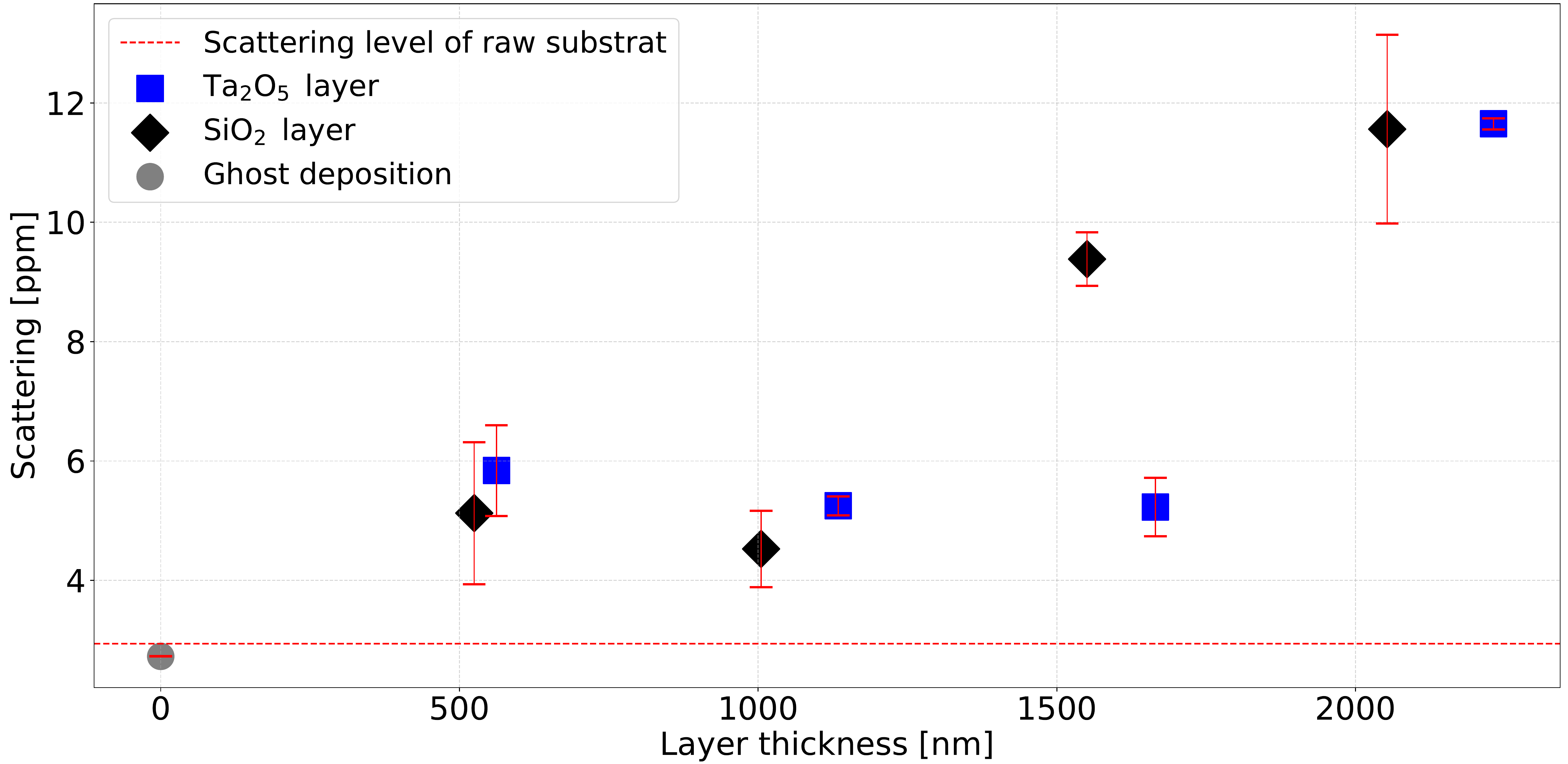} \hfill
\caption{Scattering light for the two materials, $\text{Ta}_{2}\text{O}_{5}$ and $\text{SiO}_{2}$, as a function of the layer thickness. The measurements were repeated and the markers shows the average values, while the error bars show the region containing 90\% of the results. The measured scattered light for a sample where the substrate was put in the IBS machine under the normal deposition conditions, but no deposition was performed (ghost deposition) is consistent with the value measured on raw substrate (within typical repeatability uncertainties, although the value being slightly lower can also be explained by small fluctuations in the substrate quality). }
\label{fig:diffusion}
\end{figure*}

Figure \ref{fig:diffusion} shows the scattering light of the two materials as a function of the layer thickness. $\text{Ta}_{2}\text{O}_{5}$ coating seems to have a stable scattering for the first three thinnest layers that increases for the thickest one.

To verify the constant scattered light of the lower thicknesses, a layer with a thickness negligible had be deposited on a substrate called "ghost deposition". The so-called ghost deposition refers to a sample where the substrate was put in the IBS machine under the normal deposition conditions, but no deposition was performed, and for this reason is treated as a sample with null thickness. We notice that the scattered value measured is consistent with the value measured on raw substrate (within typical repeatability uncertainties, although the value being slightly lower can also be explained by small fluctuations in the substrate quality).

\begin{table}[htbp]
\centering
\caption{\bf Scattering light and defect density measurement for each material after annealing according to their layer thickness. One Coastline substrate without coating has been measured to compare the scattering value with the different samples.}
\begin{tabular}{|l|ccc|}
\hline
 Material & Thickness & Defect density & Scattering\\
   & (nm) & (def/mm$^2$) & (ppm)\\\hline\hline
 Substrate &   0 & 0.04 & 3  \\
 \hline
           &  523 & 0.14 & 5.1 $\pm$ 1.2\\
 $\text{SiO}_{2}$ &  1005 & 0.32 & 4.5 $\pm$ 0.6\\
            &  1550 & 0.52 & 9.4 $\pm$ 0.5\\
            &  2053 & 0.61 & 11.5 $\pm$ 1.6\\\hline
            &  543 & 0.12 & 5.8 $\pm$ 0.8\\
 $\text{Ta}_{2}\text{O}_{5}$ &  1083 & 0.25 & 5.3 $\pm$ 0.2\\
           &  1626 & 1.52 & 5.2 $\pm$ 0.5\\
           &  2167 & 5.16 & 11.6 $\pm$ 0.1\\\hline
\end{tabular} 
  \label{tab:shape-functions}
\end{table}

\section{Discussion}

These experimental results show that there are several factors involved in the density of defects in the layers. The defect density increases with the deposited thickness regardless of the material.

The post-deposition annealing has well-known mechanical and optical impacts on thin-films. First it reduces the compressive strain of films coated by ion beam sputtering  thanks to a redistribution of the material in the film \cite{McLeod}. In the same time, this process induces both a decrease of the refractive index and an increase of the layer thickness that is in agreement with a decreased density of the film \cite{Granata  CQG-106402}. Moreover some structural changes have been recently highlighted in sputtered silica films that could improve the mechanical quality factor of the material \cite{Granata Materials}. Thermal treatment has clearly several effects on thin-films and the results presented in this paper indicate that it seems act as a cure for the point defects. Neverthless, the mechanism of this defect mitigation must be still studied further.

The defect density generated in the $\text{SiO}_{2}$ samples is much lower than the one of $\text{Ta}_{2}\text{O}_{5}$. However, the behavior is different for the 2 materials. For the $\text{SiO}_{2}$, the defect density is linearly dependent on the thickness, while for $\text{Ta}_{2}\text{O}_{5}$ it follows a power law. 
The hypothesis that the difference between the two materials is due to particles detaching from the walls of the chamber gradually during successive depositions is excluded since the $\text{SiO}_{2}$ deposition series was made after the deposition series of $\text{Ta}_{2}\text{O}_{5}$ without cleaning  the chamber.

However, the defect density as a function of the thickness of $\text{SiO}_{2}$ varies linearly with the thickness. Furthemore, deposition layers were not made in increasing order of thickness. Hence the defect level in the $\text{Ta}_{2}\text{O}_{5}$ layers is not due to a cleanliness issue inside the coating chamber. This behavior that follow a power law could be the consequence of processes thermally activated during the deposition. Indeed, the deposition temperature is the same for all layers but the deposition of silica layers is quite long compared to tantala. For this reason the silica substrate temperature was in a steady state during most of the deposition process, while for the tantala the temperature was possibly evolving during deposition.

Also, the relation between density of defects and the level of scattered light must be better understood. From figures \ref{fig:silice-tantale} and \ref{fig:diffusion}, we can notice that the two are not directly proportional. The scattering values would suggest that there is a kind or a size of defect which scatters more in silica than in tantala. In addition scattered light seems to be constant for the three (two) thinnest layers of $\text{Ta}_{2}\text{O}_{5}$ ($\text{SiO}_{2}$). This visible effect on both materials can be explain by the background noise of the measurement. 

The analysis of the defects size confirms that the annealing has an impact on the coating by reducing the defect size (depending on the materials) which is consistent with the scattered light measurements. 

Improving the deposition process and thermal annealing is essential if we want to reduce the amount of defects. According to our study, the use of some materials are more favourable to have defects, but this does not necessarily imply an increase in the optical scattering as we saw with the $\text{Ta}_{2}\text{O}_{5}$.

\section{Next steps}

Several items are planned so as to deal with this study in depth. 

First of all, to confirm the hypothesis that for the silica substrate temperature is in steady sate contrary to the tantala, it is planned to make several depositions with different thicknesses during which the structure would be preheated to reach thermal equilibrium so that the depositions are all made at the same temperature.

Secondly and for better understanding, we envisage to have a second set of sample with more samples per thicknesses which would allow for further statistical analyzes.

\section*{Fundings}

National Research Agency (ANR) (ANR-11-IDEX-0007);
LABEX Lyon Institute of Origins of the Université de Lyon (ANR-10-LABX-0066). The authors gratefully acknowledge the support of the European Gravitational Observatory (EGO). S. Sayah is supported by the EGO collaboration convention for the funding of a fellowship at LMA.

\section*{Acknowledgement}

We are grateful to Colin Bernet (researcher from IP2I-Lyon) and Corentin Pecontal (intern), for help and suggestions on the image recognition algorithm used to obtain the critical information on the size of the defects.

\section*{Disclosures}

The authors declare no conflicts of interest.

\end{document}